\documentclass[11pt]{article}

\usepackage{amsmath} 
\usepackage[english]{babel}
\usepackage{amsbsy}
\usepackage{amsfonts}
\usepackage{amssymb}
\usepackage{amsthm}
\usepackage[pdftex]{graphicx}
\usepackage{color}
\usepackage{float}
\usepackage{framed}
\usepackage[hyphens]{url}
\usepackage{mathtools}
\usepackage{color, colortbl}
\usepackage{xcolor}
\usepackage{mathrsfs}
\usepackage{geometry}
\usepackage[T1]{fontenc}
\usepackage[utf8]{inputenc}
\usepackage{authblk}
\usepackage{cases}
\usepackage[numbers]{natbib}
\usepackage{makecell}
\usepackage{algorithmic}
\usepackage{algorithm}
\usepackage{caption}
\usepackage{longtable}

\usepackage{subcaption}
\usepackage{ltablex}
\usepackage{wrapfig}
\usepackage{lscape}
\usepackage{rotating}
\usepackage{epstopdf}
\usepackage{appendix}

\usepackage{booktabs}
\usepackage{multirow}
\usepackage{array}

\usepackage{hyperref}
\hypersetup{
	colorlinks=true,
	linkcolor=blue,
	filecolor=blue,
	citecolor = blue,      
	urlcolor=cyan,
}

\title{\textbf{Boosted Physics-Informed Neural Networks for Solving Homogeneous and Non-Homogeneous Lane-Emden Type Equations}}
\author[]{Luke Oluwaseye Joel\thanks{oluwaseyejoel@gmail.com; ljoel@uj.ac.za}}
\affil[]{Department of Mathematics and Applied Mathematics\\
           University of Johannesburg\\ Johannesburg, South Africa}
           
 \date{}

\begin{document}

\maketitle


\begin{abstract}
The Lane-Emden Equation is a class of nonlinear differential equations commonly used to model phenomena in astrophysics and mathematical physics, such as stellar structure and thermal behavior in polytropic gas spheres. This study presents a novel hybrid framework that integrates Physics-Informed Neural Networks (PINNs) with boosting methods to enhance the accuracy and robustness of solutions for Lane-Emden type differential equations. Standard PINNs leverage physical laws encoded within the loss function, but their predictive performance can be constrained by optimization challenges and approximation errors. To address this, we introduce a post-processing boosting stage where the residual errors from the PINN prediction are modeled using some regression algorithms - including LightGBM, Random Forest, XGBoost, and Support Vector Regression (SVR). The outputs from these boosted models are aggregated to generate a corrected, high-fidelity prediction. The proposed methodology is systematically validated through quantitative metrics and comparative analysis against the standard PINN baseline. The results demonstrate that the boosted PINN framework significantly reduces prediction error, and improves generalization, offering a powerful and flexible approach for solving scientific and engineering problems governed by differential equations.
\end{abstract}


\section{Introduction}
The general non-homogeneous form of the Lane-Emden equation \cite{chandrasekhar1957, horedt2004, joel2025solving} includes an additional term $ f(\xi)$ that acts as an external source or sink:
\begin{equation}
\frac{1}{\xi^{\alpha}} \frac{d}{d \xi} \left( \xi^{\alpha} \frac{d \theta}{d \xi} \right) + \theta^n = f(\xi)
\label{eqn1}
\end{equation}
where $\alpha$ is a arbitrary constant in the radial term, $\xi$ is a dimensionless radial coordinate, $\theta$  is a dimensionless function related to the density of the star by the following equation
\begin{equation}
	\rho =\rho_{c} \theta ^{n}
\end{equation} 
for central density $ \rho _{c}$. The index $n$ is the polytropic index that appears in the polytropic equation of state, 
\begin{equation}
	P= K \rho ^{1+{\frac {1}{n}}}, 
\end{equation} 
where $P$  and $ \rho $ are the pressure and density, respectively, and $ K $ is a constant of proportionality. The term, $ f(\xi) $ represents a non-homogeneous term that can vary with the radial coordinate $ \xi $. It can be used to model external effects such as heat sources, gravitational forces, or other radial-dependent forces not accounted for by the homogeneous form.

This general form, in Equation~\eqref{eqn1}, is used in models where the system is not isolated or when external factors contribute to the dynamics. The function $ f(\xi) $ introduces additional complexity, as it can significantly affect the behavior of solutions depending on its form (e.g., constant, polynomial, or exponential in $ \xi $). Hence, the Lane-Emden equation, a type of nonlinear differential equation, plays a crucial role in astrophysics, particularly in the study of stellar structure and polytropic models of gas spheres. It describes the relationship between pressure, density, and gravitational forces in a spherically symmetric body, such as a star.

We obtain a second-order differential equation from Equation~\eqref{eqn1} by multiplying both sides of the equation by $ \xi^{\alpha} $, differentiating $ \xi^{\alpha} \frac{d \theta}{d \xi} $ using the product rule, and rewriting in a standard form. The resulting equation is as follows.
\begin{equation}
\frac{d^2 \theta}{d \xi^2} + \frac{\alpha}{\xi} \frac{d \theta}{d \xi} + \theta^n = f(\xi),
\label{eqn2}
\end{equation}
subject to initial conditions $\theta(0) = \beta_1$ and  $ \frac{d \theta}{d \xi}(0) = \beta_2 $, where $\beta_1$ and $\beta_2$  are constants. This equation represents the general form of the Lane-Emden equation with an arbitrary constant in the radial term, allowing for flexibility in modeling various radial geometries or scaling properties and a source term $ f(\xi) $.

Studies which solve different Lane-Emden type equation abound in the literature. Gurbuz and Sezer \cite{gurbuz2014laguerre} solved three examples of the Lane-Emden type equation using the laguerre polynomial approach (LPA). The results were only compared with the exact solution.  Sinha and Maroju \cite{sinha2024numerical} solved the six different Lane-Emden problems arise in isothermal gas spheres, stellar structures, and physics using the advanced variational iteration method (AVIM). The results were compared with the exact solutions and with some existing methods \cite{mall2014,el2023,umesh2021approximate,ahmad2016} in the literature. El-Essawy et al. \cite{el2023novel} solved eleven Lane-Emden type equations usimng two computational algorithms based on Monte Carlo (MC) methods. Some results of the study were compared with several existing methods \cite{wazwaz2002new,singh2018efficient,gurbuz2014laguerre,verma2019,mall2014,parand2017numerical,parand2010approximation,dizicheh2020novel,wazwaz2001new} in the literature. 

Mall and Chakraverty \cite{mall2014} solved seven examples of Lane-Emden type equation using Chebyshev Neural Network (ChNN) model and compared the results with only the exact solution. Verma and Kumar \cite{verma2019} solved three examples of Lane-Emden type differential equation using multi-layer perceptron artificial neural network (MLP ANN) technique and the results were compared with the exact solutions and the results in \cite{mall2014}. Ahmad et al. \cite{ahmad2016} solved the three cases of the Lane-Emden type equation with five bio‑inspired computational heuristic - pattern search(PS), genetic algorithm (GA), sequential quadratic programming (SQP), hybrid pattern search with sequential quadratic programming (PS-SQP) and  hybrid genetic algorithm with sequential quadratic programming (GA-SQP). The results from these algorithms were compared with the exact solution and the ChNN method \cite{mall2014}. The study also performed some statistical comparison of the five heuristic algorithms used.

Singh et al. \cite{singh2018reliable}, in their study, solved ten examples of the Lane-Emden type equations using an algorithm based on the operational matrix of integration for Jacobi polynomials and the collocation method. The results of the study was shown for five different cases of jacobi polynomials - the Legendre polynomial (P1), the Chebyshev polynomials of the second kind (P2), the Chebyshev polynomials of the third kind (P3), the Chebyshev polynomials of the fourth kind (P4), and the Gegenbauer (or ultraspherical) polynomials (P5). These results were compared to the exact solution and some existing methods \cite{parand2010approximation,pandey2012solution,pandey2012soln,dehghan2013technique,cuaruntu2013approximate} in the literature. 

This study introduces the integration of Physics-Informed Neural Networks (PINN) with ensemble boosting methods (LightGBM, Random Forest, XGBoost, SVR) for residual error correction in solving differential equations. It proposes a decoupled approach where PINN provides a physics-constrained baseline solution and ML models correct residual errors without altering the physics formulation. We showcase this by solving six different examples of Lane-Emden type equations - both homogeneous and non-homogenous.

The reminder of the paper is divided into the following parts.  In Section \ref{secpinns}, we discuss the PINN methods that is used in this study, the hyperparameter tuning that was done using the Bayesian optimization in Optuna framework and the machine learning (ML) algorithms used for boosting the PINN method. Section \ref{secresults} details the six different problems we solve using PINN and their comparative solutions with some existing methods in the literature. The section also showcase the results for the Boosted PINNs. And Section \ref{secconclusion} is the conclusion of the study.

\section{Boosted Physics Informed Neural Networks}
\label{secpinns}
Physics-informed neural networks (PINNs) \cite{raissi2019, karniadakis2021} are simply artificial neural networks that are trained to solve any supervised learning tasks while respecting the given laws of physics described by the given ordinary or partial differential equations \cite{joel2025solving}. To make the model aware of these underlying equations, the first step is to define a loss function that captures the constraints imposed by the physics. The PINNs is then trained to directly approximate the solution to the differential equation. That is,  
\begin{equation}
	\theta_{PINN}(\gamma) \approx \theta(\xi)
\end{equation}
where $\gamma$ are the free parameters of the PINNs, and $\theta(\xi)$ is the exact solution. The loss function is given as:
	\begin{equation}
		Loss = L_{equation}  + L_{IC1} + L_{IC2}
		\label{lossEq}
	\end{equation}
	where $L_{equation}$, $L_{IC1}$ and $L_{IC2}$ are the loss for the differential equation, the initial and boundary conditions respectively. Each of these terms is as follows.
	\begin{equation}
		L_{equation} = \frac{1}{N} \sum_{i=1}^{N} \left[ \frac{d^2 \theta_{PINN}}{d \xi^2}|_{\xi_i} + \frac{\alpha}{\xi} \frac{d \theta_{PINN}}{d\xi}|_{\xi_i} + \theta^n_{PINN}(\xi_i)  -   f_{PINN}(\xi)  \right]^2
	\end{equation}
	
	\begin{equation}
		L_{IC1} =  \left ( \theta_{PINN} (\xi_i) - \beta_1 \right )^2
	\end{equation}
	
	\begin{equation}
		L_{IC2} =  \left ( \frac{d \theta_{PINN}}{d\xi}|_{\xi_i} - \beta_2 \right )^2.
	\end{equation}

For the PINN solutions of the various Lane-Emden equation, we build a fully connected neural network, similarly to the one in \cite{joel2025solving}, with the number of hidden layers, number of neurons, activation function and learning rate suitable for each problem through the hyperparameter tuning using Bayesian Optimization.


The optimal hyperparameter tuning obtained using Optuna framework is given in Table~\ref{hyParam}. On the advantages of using PINNs, PINNs are adept at learning from both boundary data and internal equation constraints, making them ideal for nonlinear and complex equations. Also, they do not require discretization, thus they can better handle singularities at $x=0$ in the Lane-Emden equations. We note that in the solution of Equation~\eqref{eqn2} using PINNs, we do not multiply $\xi$ across the equation. We solved the eqution with the singularity included. 
\begin{table}[H]
	\centering
	\caption{Hyperparameter Tuning for the Problems}
	\begin{tabular}{|c|c|c|c|c|c|c|c|}
		\hline
		Items & \makecell{Options} & Prob1  & Prob2 & Prob3 & Prob4 & Prob5 & Prob6 \\
		\hline
		\textit{\makecell{Number \\ of \\ hidden\\  layers}} & 2, 3, 4 &  3 & 2 & 4  & 3 & 4 & 4 \\
		\hline
		\textit{\makecell{Number \\ of \\ neurons \\ per layer}} &  4, 8, 16, 32, 64 & 32  & 64 & 64 & 16 & 16 &  32 \\
		\hline
		\textit{\makecell{Activation \\ Function}} & \makecell{ReLU, \\ Tanh, \\ SiLU,  \\ Sigmoid, \\ Elu, \\ Softplus} & Tanh  & Sigmoid  & SiLU & Softplus & Sigmoid & Sigmoid \\
		\hline 
		\textit{Optimizer} & \makecell{Adam, \\ Adagrad,  \\ SGD, \\  RMSprop, \\ AdamW, \\ Adamax,  \\ NAdam} &  Adamax & Adamax & Adamax  & Adam & AdamW  & Adam \\
		\hline
		\textit{\makecell{Learning \\ rate}} & \makecell{0.01, \\ 0.001, \\ 0.0001, \\ 0.0001} & 0.001 & 0.01 & 0.01 & 0.01 &  0.001 &  0.001 \\
		\hline	
	\end{tabular}
	\label{hyParam}
\end{table}	

\begin{figure}[t!]
	\centering
	\includegraphics[height=3.0in]{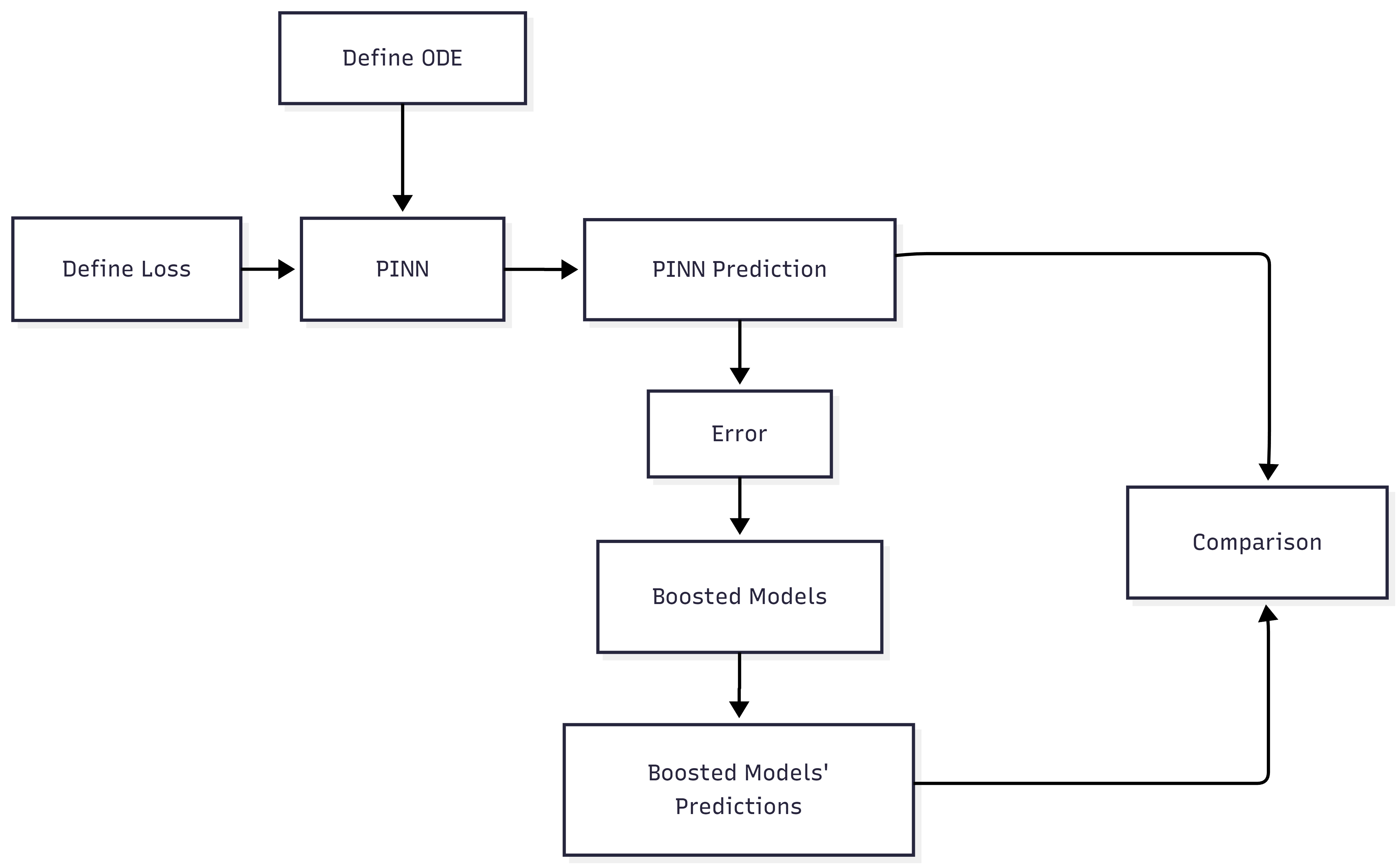}
	\caption{Boosted PINNs Flowchart}
	\label{boostedFig}
\end{figure}

Algorithm~\ref{alg:boosted_pinn} and the flowchart in Figure \ref{boostedFig} highlight the steps taken in the boosted PINNs. The Boosted PINN algorithm presented in Algorithm~\ref{alg:boosted_pinn} operates through a systematic two-stage framework that combines physics-based learning with data-driven error correction. The algorithm begins by defining a physics-informed loss function $\mathcal{L}_{\text{PINN}} = \mathcal{L}_{\text{res}} + \lambda \mathcal{L}_{\text{bc}}$, where $\mathcal{L}_{\text{res}}$ enforces the governing differential equation $F(x, y, y', y'')=0$ and $\mathcal{L}_{\text{bc}}$ enforces boundary conditions $B(x, y, y')=0$ with weighting parameter $\lambda$. The PINN is trained by solving the optimization problem $\theta^\ast = \arg\min_{\theta} \mathcal{L}_{\text{PINN}}(\theta)$ to obtain parameters $\theta^\ast$, yielding an initial solution $\hat{y}_{\text{PINN}}(x)$. The residual error $r(x) = y_{\text{exact}}(x) - \hat{y}_{\text{PINN}}(x)$ is then computed across the test domain $\Omega_{\text{test}}$. In the boosting stage, multiple machine learning models $M_j \in \mathcal{M}$ (including LightGBM, Random Forest, XGBoost, and SVR) are independently trained to predict this residual error, with each model learning the mapping from spatial coordinates $x$ to error values $r(x)$. Each trained model produces a correction term $\hat{r}_j(x) = M_j(x)$, which is added to the original PINN prediction to generate boosted solutions $\hat{y}_{\text{boosted}, j}(x) = \hat{y}_{\text{PINN}}(x) + \hat{r}_j(x)$. Finally, performance metrics including absolute error (AE), mean absolute error (MAE), and maximum error are computed for each boosted model, enabling comparative evaluation and selection of the optimal error-corrected solution. This hybrid approach effectively decouples physics-based approximation from data-driven refinement, allowing systematic error reduction while maintaining the interpretable structure of the physics-informed formulation. In Table~\ref{algorithms_summary},  we present a summary of the Machine learning algorithms used for boosting the PINN method.

	\begin{table}[htbp]
		\centering
		\caption{Summary of Machine Learning Algorithms Used for Boosting PINN}
		\label{algorithms_summary}
		\begin{tabular}{|p{4cm}|p{3cm}|p{7cm}|}
			\hline
			\textbf{Algorithm} & \textbf{Type}  & \textbf{Role in Boosted PINN Framework} \\ \hline
			Random Forest (RF) & Ensemble (Bagging)  &
			Captures non-linear patterns and interactions in PINN residuals through decorrelated decision trees, reducing variance and mitigating overfitting. \\ \hline
			Extreme Gradient Boosting (XGBoost) & Ensemble (Boosting)  &
			Sequentially corrects PINN errors with regularization, using second-order gradients for precise residual modeling while controlling model complexity. \\ \hline
			Light Gradient Boosting Machine (LightGBM) & Ensemble (Boosting) &
			Efficiently models PINN residuals using histogram-based algorithms and leaf-wise growth, enabling fast training on error datasets with many sample points. \\ \hline
			Support Vector Regression (SVR) & Kernel Method &
			Provides smooth interpolation of PINN error surfaces using $\epsilon$-insensitive loss and kernel functions, effective for capturing global error trends. \\ \hline
		\end{tabular}
	\end{table}

\begin{algorithm}[H]
	\caption{Boosted Physics-Informed Neural Network}
	\label{alg:boosted_pinn}
	\begin{algorithmic}[1]
		\REQUIRE Differential equation $F(x, y, y', y'')=0$,
		boundary conditions $B(x, y, y')=0$,
		training domain $\Omega=\{x_i\}$,
		boosting models $\mathcal{M}=\{M_1, M_2, M_3, M_4\}$.
		\ENSURE Boosted solution $\hat{y}_{\text{boosted}}(x)$.	
		\STATE Define the physics-informed loss
		 $ 	\mathcal{L}_{\text{PINN}} = \mathcal{L}_{\text{res}} + \lambda \mathcal{L}_{\text{bc}}, $
		where $\mathcal{L}_{\text{res}}$ enforces $F(x,y,y',\ldots)=0$ and
		$\mathcal{L}_{\text{bc}}$ enforces $B(x,y,y',\ldots)=0$.
		\STATE Train the PINN by solving 
		$\theta^\ast = \arg\min_{\theta} \mathcal{L}_{\text{PINN}}(\theta). $
		\STATE Predict the PINN solution on a test grid $\Omega_{\text{test}}$:
		$ \hat{y}_{\text{PINN}}(x) = \text{PINN}(x). $	
		\STATE Compute the residual error $
		r(x)= y_{\text{exact}}(x) - \hat{y}_{\text{PINN}}(x). $	
		\FOR{each boosting model $M_j \in \mathcal{M}$}
		\STATE Train $M_j$ using inputs $x \in \Omega_{\text{test}}$ and targets $r(x)$.
		\STATE Obtain the predicted correction:
		$ 	\hat{r}_j(x) = M_j(x). $
		\STATE Form the boosted prediction:
		$ \hat{y}_{\text{boosted}, j}(x)
		= \hat{y}_{\text{PINN}}(x) + \hat{r}_j(x). $
		\ENDFOR		
		\FOR{each boosted model $j$}
		\STATE Compute performance metrics:
		$ \text{AE}_j,\;\text{MAE}_j,\;\text{MaxError}_j. $
		\ENDFOR		
		\STATE Select the best boosted model or report all boosted solutions.		
		\RETURN Original PINN solution $\hat{y}_{\text{PINN}}(x)$,
		boosted solutions $\{\hat{y}_{\text{boosted},j}(x)\}$,
		and corresponding performance metrics.		
	\end{algorithmic}
\end{algorithm}

\section{Results}
\label{secresults}
The following are the problems we consider, they are picked from the following literature \cite{mall2014, mall2015b, verma2019, ahmad2016, abdelhakem2021two, el2023novel}. We note that the $x$ at the denominator of the term $\frac{\alpha}{x}$, in each problem was not multiplied across the equation, in our case, as most of the studies used for comparison did. This could affect the order of PINN absolute error, yet the PINN results are still comparable to the other numerical methods and in some cases achieve lower absolute errors.

\paragraph{Problem 1:}
The first example is given as 
\begin{equation}
	\frac{d^2 y}{d x^2} + \frac{2}{x} \frac{d y}{dx} + y^5 = 0
	\label{prob1}
\end{equation}
with the initial conditions $ y(0) = 1,  \frac{d y}{d x}(0) = 0 $. The exact solution is 
\begin{equation}
	y(x) = \left ( 1 + \frac{x^2}{3} \right )^{-\frac{1}{2}}.
\end{equation}

\begin{figure}[t!]
		\centering
		\begin{subfigure}[t]{0.45\textwidth}
			\centering
			\includegraphics[height=1.8in]{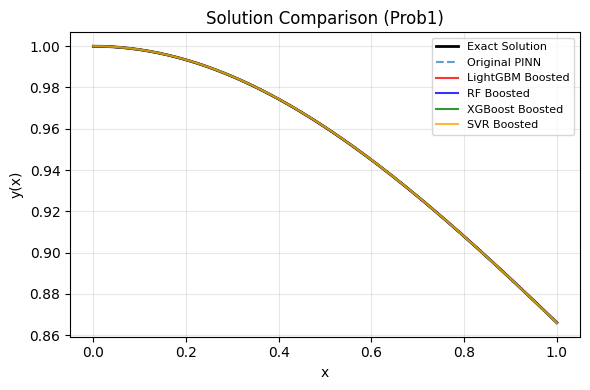}
			\caption{Solution for 30 000 iterations}
			\label{n1v11}
		\end{subfigure}%
		\begin{subfigure}[t]{0.5\textwidth}
			\centering
			\includegraphics[height=1.8in]{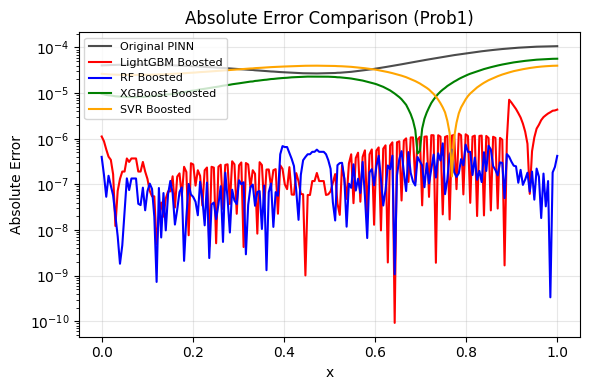}
			\caption{The Absolute Error}
			\label{n1v12}
		\end{subfigure}
		\caption{The Solutions and Absolute Error for Problem 1}
		\label{fig1a}
	\end{figure}

\begin{figure}[t!]
	\centering
	\begin{subfigure}[t]{0.45\textwidth}
		\centering
		\includegraphics[height=1.8in]{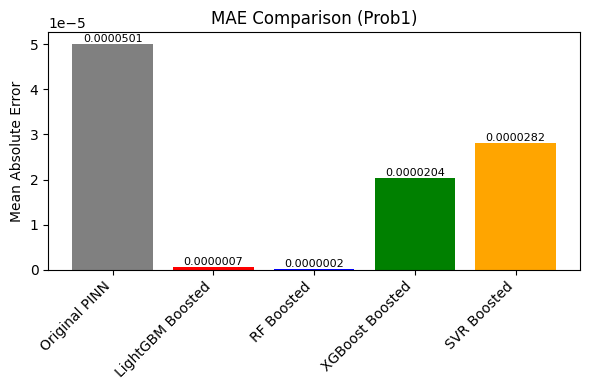}
		\caption{Mean Absolute Error Comparison}
		\label{n1v21}
	\end{subfigure}%
	\begin{subfigure}[t]{0.5\textwidth}
		\centering
		\includegraphics[height=1.8in]{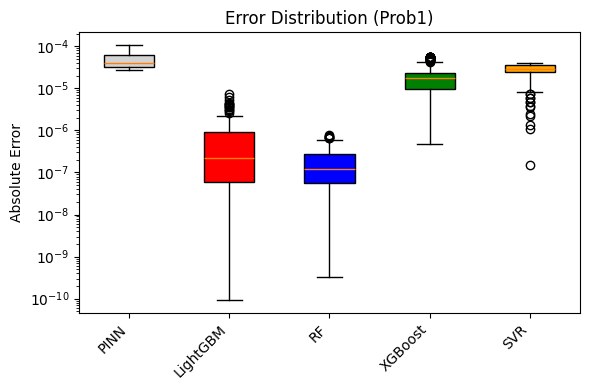}
		\caption{The Absolute Error Distribution}
		\label{n1v22}
	\end{subfigure}
	\caption{The Mean Absolute Error and Absolute Error Distribution for Problem 1}
	\label{fig1b}
\end{figure}

Figure \ref{n1v11} shows the solutions obtained by the PINN and the Boosted PINNs in comaprison to the exact solution while Figure \ref{n1v12} shows the absolute error for PINN and ML-Boosted PINNs. Figures \ref{n1v21} and \ref{n1v22} show the mean absolute error and the error distribution for PINN and boosted PINNs respectively. RF has the lowest error, followed by LightGBM, then XGBoost and lastly SVR. Table \ref{comp1} presents the comparison of the absolute error of PINN with the different methods in the literature - Multilayer Perceptron Artificial Neural Network (MLP ANN) \cite{verma2019}, Chebyshev Neural Network (ChNN) \cite{mall2014}, Particle Swarm with Sequential Quadratic Programming (PS-SQP) \cite{ahmad2016} and Genetic Algorithm with SQP (GA-SQP) \cite{ahmad2016} -  for solving Problem 1 at selected values of $x \in [0.1, 1.0]$. The results show that PINN consistently achieves the lowest or among the lowest errors across almost the entire domain. Its errors remain stable in the range $ \sim 10^{-5}$, while the errors of other methods, particularly PS‑SQP and GA‑SQP, increase by orders of magnitude as $x \rightarrow 1$. This suggests that the PINN offers superior accuracy and robustness for this problem, maintaining low error uniformly without the degradation seen in traditional and hybrid optimization methods.

Table \ref{comp1b} compares the absolute error of a standard PINN with the four ML‑boosted algorithms for solving Problem 1. The boosted methods - especially LightGBM‑Boosted and RF‑Boosted-  consistently achieve significantly lower errors (ranging from $\sim 10^{-8}$ to $10^{-6}$) compared to the standard PINN ($10^{-5}$) with RF‑Boosted often yielding the smallest errors. XGBoost‑Boosted and SVR‑Boosted also generally outperform the standard PINN, though not as dramatically as the first two. This indicates that boosting techniques can greatly enhance accuracy for solving a differential equation.  
	
\begin{table}[H]
		\centering
		\caption{Absolute Error Comparison for Problem 1}
		\begin{tabular}{|c|c|c|c|c|c|}
			\hline
			x & PINN & MLP ANN \cite{verma2019} & ChNN \cite{mall2014} & PS-SQP \cite{ahmad2016}   &  GA-SQP \cite{ahmad2016} \\
			\hline
			0.1 & 4.17e-05  & 1.54e-04 & 2.00e-04  &  6.25e-05 & 9.28e-06 \\
			\hline
			0.2 &  3.88e-05 &  4.22e-04 &  1.00e-04 &  1.06e-04 & 7.54e-05 \\
			\hline
			0.3 & 3.37e-05    &  6.04e-04 & 4.60e-03 &  1.13e-04 & 3.38e-04 \\
			\hline
			0.4 &  2.86e-05   & 6.03e-04 &  3.20e-03 &  8.25e-04  & 1.02e-03 \\
			\hline	
			0.5 & 2.74e-05  & 4.12e-04 &  7.60e-03 & 2.28e-03 &  2.41e-03 \\
			\hline
			0.6 & 3.45e-05  &  9.20e-05 &  3.80e-03 &  4.77e-03 & 4.97e-03 \\
			\hline
			0.7 & 5.18e-05  &  2.54e-04  &  3.20e-03 &  8.67e-03 & 8.88e-03 \\
			\hline
			0.8 & 7.55e-05   &  5.00e-04  &  2.00e-04 & 1.43e-02  &  1.45e-02 \\
			\hline
			0.9 & 9.72e-05   &  5.10e-04  &  4.40e-03  &  2.22e-02 &  2.23e-02 \\
			\hline
			1.0 & 1.06e-04  & 1.59e-04 &  9.00e-04  & 3.25e-02 & 3.69e-02  \\
			\hline
		\end{tabular}
		\label{comp1}
	\end{table}

		\begin{table}[H]
		\centering
		\caption{Absolute Error Comparison of PINNs with the Boosted Algorithms - Problem 1}
		\begin{tabular}{|c|c|c|c|c|c|}
			\hline
			x & PINN  & LightGBM-Boosted & RF-Boosted & XGBoost-Boosted   &  SVR-Boosted \\
			\hline
			0.1 &  4.17e-05 &  1.31e-07	& 7.11e-08	& 8.40e-06	& 2.52e-05 \\ \hline
			0.2 & 3.88e-05  & 2.73e-07  & 5.29e-08 & 1.13e-05 &	2.81e-05 \\ \hline	
			0.3 &  3.37e-05   &   2.61e-07 & 1.23e-07 &	1.64e-05 &	3.32e-05 \\ \hline
			0.4 &  2.86e-05   &  1.00e-07   & 	6.58e-07 &	2.15e-05	 & 	3.83e-05 \\ \hline
			0.5 &   2.74e-05   &  7.06e-08  &  2.19e-07	 & 	2.28e-05 &	3.95e-05 \\ \hline
			0.6 &  3.45e-05 &  5.49e-07	& 2.46e-07 & 1.57e-05 & 3.25e-05 \\ \hline
			0.7 &  5.18e-05 &  3.47e-08	&  2.61e-07	& 1.67e-06	& 1.51e-05 \\ \hline
			0.8 & 7.55e-05 & 1.15e-06  & 5.16e-07 & 2.54e-05  & 8.61e-06 \\ \hline
			0.9 & 9.72e-05   &  5.25e-06 & 2.57e-07	& 4.71e-05	& 3.03e-05	\\ \hline
			1.0 & 1.06e-04  &  4.35e-06	& 4.21e-07  & 5.67e-05	& 3.99e-05	\\	\hline
		\end{tabular}
		\label{comp1b}
	\end{table}

\paragraph{Problem 2:}
The second example is
\begin{equation}
	\frac{d^2 y}{d x^2} + \frac{2}{x} \frac{d y}{dx} - 2(2x^2 + 3)y = 0
	\label{prob2}
\end{equation}
with the initial conditions $ y(0) = 1, \frac{d y}{d x}(0) = 0 $.  The exact solution is 
\begin{equation}
	y(x) = e^{x^2}.
\end{equation} 

\begin{figure}[t!]
	\centering
	\begin{subfigure}[t]{0.45\textwidth}
		\centering
		\includegraphics[height=1.8in]{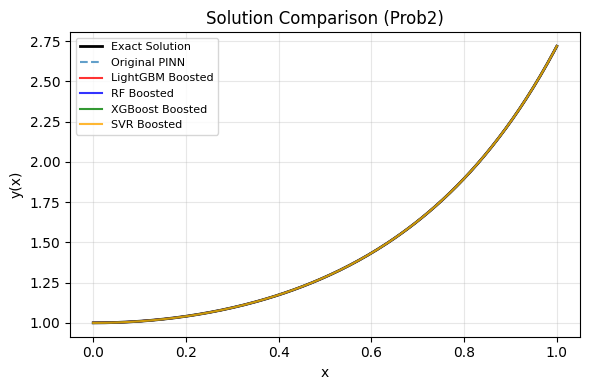}
		\caption{Solution for 50 000 iterations}
		\label{n2v11}
	\end{subfigure}%
	~ 
	\begin{subfigure}[t]{0.5\textwidth}
		\centering
		\includegraphics[height=1.8in]{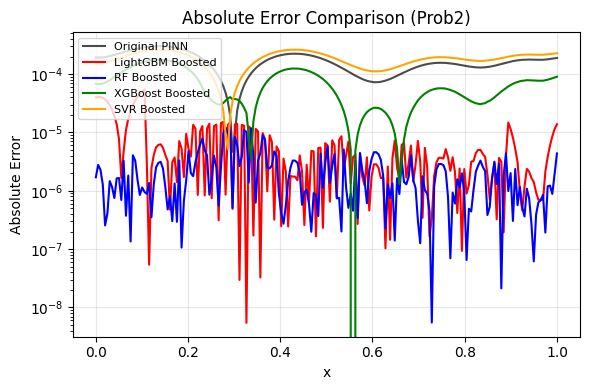}
		\caption{The Absolute Error}
		\label{n2v12}
	\end{subfigure}
	\caption{The Solutions and Absolute Error for Problem 2}
	\label{fig2a}
\end{figure}

\begin{figure}[t!]
	\centering
	\begin{subfigure}[t]{0.45\textwidth}
		\centering
		\includegraphics[height=1.8in]{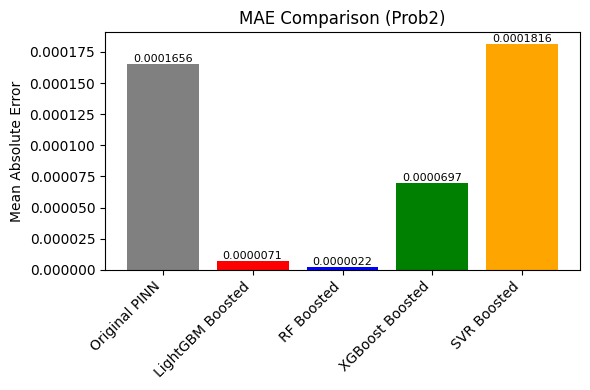}
		\caption{The Mean Absolute Error Comparison}
		\label{n2v21}
	\end{subfigure}%
	~ 
	\begin{subfigure}[t]{0.5\textwidth}
		\centering
		\includegraphics[height=1.8in]{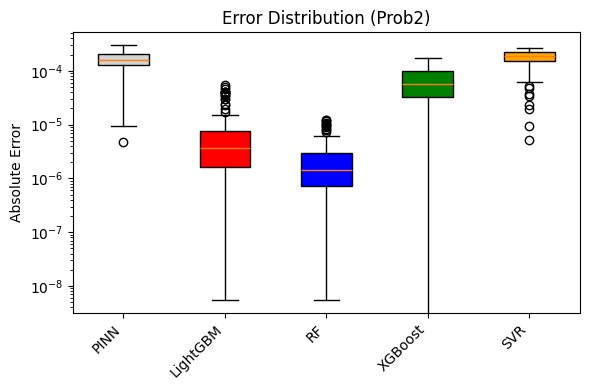}
		\caption{The Absolute Error Distribution}
		\label{n2v22}
	\end{subfigure}
	\caption{The Mean Absolute Error and Absolute Error Distribution for Problem 2}
	\label{fig2b}
\end{figure}

For problem 2, Figure~\ref{n2v11} shows the solutions as compared to the exact solution while Figure~\ref{n2v12} shows the absolute errors. Figures \ref{n2v21} and \ref{n2v22} show the mean absolute error and the error distribution for PINN and boosted PINNs respectively. RF has the lowest MAE and error distribution, followed by LightGBM. The results in Table \ref{comp2} compare the absolute errors of PINN with four other numerical methods for solving Problem 2. The PINN demonstrates competitive and often superior performance, particularly when compared to the Chebyshev Neural Network (ChNN) \cite{mall2014}, and the two Monte Carlo-based methods - $MC_1$ and $MC_2$ \cite{el2023novel}. Notably, the PINN's errors remain stable within the order of \(10^{-4}\) to \(10^{-6}\), while the errors for \(MC_1\) and \(MC_2\) tend to increase as \(x\) approaches 1. 

The results in Table \ref{comp2b} compare the absolute errors of a standard PINN with the four boosted ML algorithms for solving Problem 2. In this comparison, the boosted methods - especially LightGBM-Boosted and RF-Boosted - consistently and significantly outperform the standard PINN across the entire domain, often by one to three orders of magnitude. For instance, the PINN's errors range from approximately \(10^{-4}\) to \(10^{-6}\), while RF-Boosted achieves errors as low as \(10^{-7}\) to \(10^{-8}\) at several points. XGBoost-Boosted generally yield lower errors than the PINN, though its advantage is less pronounced than that of the tree-based boosting algorithms. However, SVR-Boosted seems not to improved the error of PINN. This result demonstrates that applying ensemble boosting techniques can substantially enhance the accuracy of the solution of PINN.

\begin{table}[H]
	\centering
	\caption{Absolute Error Comparison for Problem 2}
	\begin{tabular}{|c|c|c|c|c|c|}
		\hline
		x & PINNs  & MLP ANN \cite{verma2019} & ChNN \cite{mall2014} & $MC_1$ \cite{el2023novel} & $MC_2$ \cite{el2023novel} \\
		\hline
		0.1 &  2.83e-04  &  2.58e-04 & 7.00e-03 &  1.99e-04  &  1.28e-05 \\
		\hline
		0.2 &  2.45e-04  &  2.67e-04 &  1.30e-02 &   4.13e-04   &  6.16e-04 \\
		\hline
		0.3 & 9.54e-06   &  7.00e-05 &  3.00e-03 &  6.53e-04   &  7.64e-04   \\
		\hline
		0.4 & 2.11e-04   & 3.00e-05 & 1.34e-01 &  9.35e-04  &  2.19e-06  \\
		\hline	
		0.5 & 1.68e-04  & 3.00e-06 & 2.60e-03 &  1.28e-03  &  1.65e-03  \\
		\hline
		0.6 &  7.26e-05   & 2.00e-05 & 2.10e-03  &  1.71e-03  &  3.36e-03  \\
		\hline
		0.7 &  1.38e-04   &  7.00e-05 &  8.50e-03  &  2.28e-03  &  3.11e-03  \\
		\hline
		0.8 &  1.37e-04  &  1.15e-04 &  4.10e-03  &  3.03e-03 &  1.79e-03  \\
		\hline
		0.9 &  1.67e-04  &   1.28e-04  & 8.70e-03 &  4.04e-03  &  4.01e-04  \\
		\hline
		1.0 &  1.89e-04  &   4.70e-05 & 2.97e-02 &  5.43e-03  &  3.60e-03 \\
		\hline
	\end{tabular}
	\label{comp2}
\end{table}

\begin{table}[H]
	\centering
	\caption{Absolute Error Comparison of PINNs with the Boosted Algorithms - Problem 2}
	\begin{tabular}{|c|c|c|c|c|c|}
		\hline
		x & PINN  & LightGBM-Boosted & RF-Boosted & XGBoost-Boosted   &  SVR-Boosted \\
		\hline
		0.1 & 2.83e-04 & 5.04e-05 & 1.14e-06 & 1.58e-04 & 2.44e-04 \\ \hline
		0.2 & 2.45e-04  &  5.21e-06	& 4.77e-06 & 1.21e-04 & 2.06e-04 \\ \hline	
		0.3 & 9.54e-06  &  1.37e-05	 & 	8.38e-06 & 	3.74e-05 &	4.82e-05 \\ \hline
		0.4 &  2.11e-04  &  2.46e-07 & 	4.19e-07 & 1.12e-04 & 2.50e-04 \\ \hline
		0.5 &  1.68e-04  &  6.03e-06	& 5.78e-06	& 6.95e-05	& 2.07e-04 \\ \hline	
		0.6 &  7.26e-05  &  2.84e-06 &	4.58e-06 & 2.65e-05	& 1.11e-04 \\ \hline
		0.7 &  1.38e-04  &   3.11e-06 & 1.26e-07 & 3.93e-05 & 1.77e-04 \\ \hline		
	    0.8 &  1.37e-04 &  8.95e-07 & 6.52e-08 & 3.81e-05 & 1.75e-04 \\ \hline
	    0.9 & 1.67e-04  &  8.12e-06 & 3.04e-07 & 6.82e-05 &	2.05e-04 \\ \hline		    
	    1.0 & 1.89e-04 & 1.38e-05 & 4.36e-06 & 9.01e-05 & 2.27e-04	\\ \hline    
	\end{tabular}
	\label{comp2b}
\end{table}

\paragraph{Problem 3:}
The third example is given as follows.
\begin{equation}
	\frac{d^2 y}{d x^2} + \frac{2}{x} \frac{d y}{dx} + y = 6 + 12x + x^2 + x^3, \quad 0 \leq x \leq 1
	\label{prob3}
\end{equation}
with the initial conditions $ y(0) = 0, \frac{d y}{d x}(0) = 0 $. The exact solution is 
\begin{equation}
	y(x) = x^2 + x^3.
\end{equation}

\begin{figure}[t!]
\centering
\begin{subfigure}[t]{0.45\textwidth}
	\centering
	\includegraphics[height=1.8in]{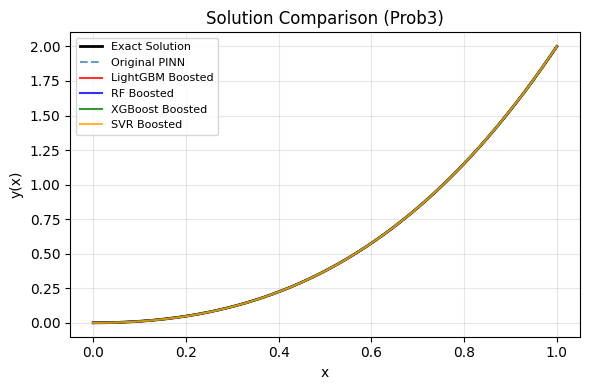}
	\caption{Solution for 30 000 iterations}
	\label{n3v11}
\end{subfigure}%
~ 
\begin{subfigure}[t]{0.5\textwidth}
	\centering
	\includegraphics[height=1.8in]{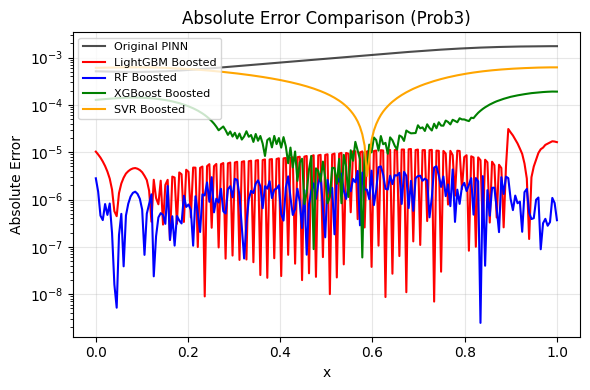}
	\caption{The Absolute Error}
	\label{n3v12}
\end{subfigure}
\caption{The Solutions and Absolute Error for Problem 3}
\label{fig3a}
\end{figure}

\begin{figure}[t!]
	\centering
	\begin{subfigure}[t]{0.45\textwidth}
		\centering
		\includegraphics[height=1.8in]{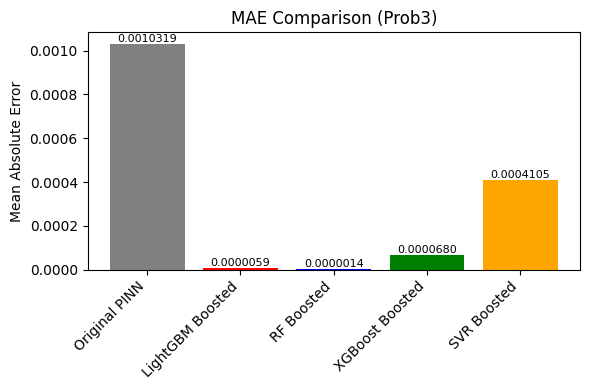}
		\caption{The Mean Absolute Error}
		\label{n3v21}
	\end{subfigure}
	~ 
	\begin{subfigure}[t]{0.5\textwidth}
		\centering
		\includegraphics[height=1.8in]{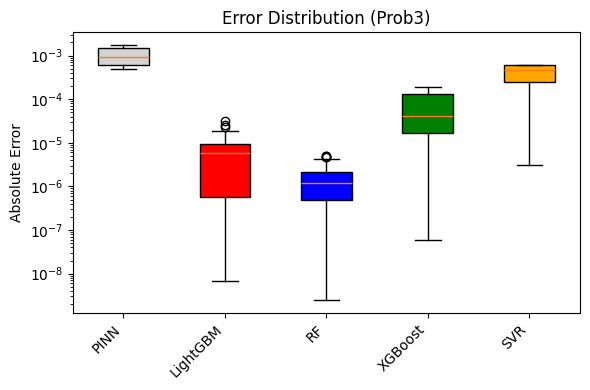}
		\caption{The Absolute Error Distribution}
		\label{n3v22}
	\end{subfigure}
	\caption{The Mean Absolute Error and Absolute Error Distribution for Problem 3}
	\label{fig3b}
\end{figure}

The solution and absolute error for the third problem is shown in Figures \ref{n3v11} and \ref{n3v12}. Also, Figures \ref{n3v21} and \ref{n3v22} show the mean absolute error and the error distribution for PINN and boosted PINNs respectively. RF has the lowest MAE and error distribution, followed by LightGBM. Table \ref{comp3} compares the absolute errors of PINN with three other methods - ChNN \cite{mall2014} and two Monte Carlo variants (\(MC_1\), \(MC_2\)) \cite{el2023novel} - for solving Problem 3. The PINN generally demonstrates more stable and moderate error growth across the domain compared to the ChNN. While the \(MC_2\) method yields the smallest errors, particularly in the latter half of the domain for this problem, the PINN consistently outperforms \(MC_1\). 

The results in Table \ref{comp3b} compare the absolute errors of a standard PINN with four boosted ML models for Problem 3. In this comparison, the boosted algorithms, particularly LightGBM-Boosted and RF-Boosted, consistently outperform the standard PINN, achieving errors that are several orders of magnitude smaller (ranging from $10^{-6}$ to $10^{-8}$) across nearly all sampled points. The XGBoost-Boosted method also shows a significant improvement over the PINN, though its errors are generally higher than those of LightGBM and RF. The SVR-Boosted model, while sometimes better than the PINN, is less accurate and consistent than the tree-based boosting methods.

\begin{table}[H]
	\centering
\caption{Absolute Error Comparison for Problem 3}
\begin{tabular}{|c|c|c|c|c|}
	\hline
	Input data & PINNs  &  ChNN \cite{mall2014} &  $MC_1$ \cite{el2023novel} & $MC_2$ \cite{el2023novel}  \\
	\hline
	0.1 & 4.98e-04  &  7.0e-04 & 2.22e-04  &  1.47e-04  \\
	\hline
	0.2 & 5.46e-04   &   1.0e-03 &  4.96e-04   &  2.99e-04 \\
	\hline
	0.3 & 6.51e-04   &  6.0e-04  & 8.19e-04   &  5.01e-04  \\
	\hline
	0.4 & 7.86e-04   &  2.6e-03 &  1.19e-03  &  5.96e-04 \\
	\hline	
	0.5 & 9.49e-04    &  7.4e-03 & 1.60e-03  &  6.55e-04   \\
	\hline
	0.6 &  1.14e-03   &   4.8e-03  & 2.07e-03  &  8.69e-04  \\
	\hline
	0.7 & 1.37e-03  &  3.3e-03  &   2.58e-03  &  5.73e-04 \\
	\hline
	0.8 & 1.57e-03   &   8.7e-03 &  3.13e-03 &  1.17e-04 \\
	\hline
	0.9 & 1.69e-03   &  1.0e-04  &  3.73e-03  &  6.87e-04  \\
	\hline
	1.0 &  1.73e-03   &  3.6e-03 &  4.36e-03  &  1.66e-03 \\
	\hline
\end{tabular}
\label{comp3}
\end{table}

\begin{table}[H]
	\centering
	\caption{Absolute Error Comparison of PINNs with the Boosted Algorithms - Problem 3}
	\begin{tabular}{|c|c|c|c|c|c|}
		\hline
		x & PINN  & LightGBM-Boosted & RF-Boosted & XGBoost-Boosted   &  SVR-Boosted \\
		\hline
		0.1 & 4.98e-04 & 3.77e-06 & 6.13e-07 & 1.41e-04 &  6.20e-04 \\ \hline
		0.2 & 5.46e-04  &  3.84e-06 & 7.85e-07 & 9.32e-05 & 5.71e-04 \\ \hline	
		0.3 & 6.51e-04 &  6.05e-06 & 2.73e-06 & 2.57e-05 & 4.66e-04 \\ \hline
		0.4 &  7.86e-04 & 2.41e-08 & 4.90e-07 & 1.25e-05	 & 	3.31e-04 \\ \hline
		0.5 & 9.49e-04 &  9.04e-06 & 2.59e-06 &  8.34e-07 & 1.68e-04	
		 \\ \hline	
		0.6 & 1.14e-03 &  1.06e-05 & 7.63e-07 & 2.04e-05 & 	2.86e-05 \\ \hline
		0.7 & 1.37e-03 & 1.10e-07 & 	3.18e-06 & 3.19e-05	& 2.52e-04 \\ \hline		
		0.8 & 1.57e-03  & 8.62e-06 &	1.51e-06 & 4.71e-05 & 4.57e-04  \\ \hline
		0.9 & 1.69e-03 & 2.26e-05 & 5.96e-07 &  1.51e-04 & 5.81e-04 \\ \hline		    
		1.0 & 1.73e-03  & 1.63e-05 & 3.65e-07 &  1.90e-04	&  6.20e-04 	\\ \hline    
	\end{tabular}
	\label{comp3b}
\end{table}

\paragraph{Problem 4:}
The fourth example is 
\begin{equation}
	\frac{d^2 y}{d x^2} + \frac{8}{x} \frac{d y}{dx} + xy = x^5 - x^4 + 44x^2 - 30x, \quad x \geq 0
	\label{prob4}
\end{equation}
with the initial conditions $ y(0) = 0, \frac{d y}{d x}(0) = 0$. The exact solution is  
\begin{equation}
	y(x) = x^4 - x^3.
\end{equation}

\begin{figure}[t!]
\centering
\begin{subfigure}[t]{0.45\textwidth}
	\centering
	\includegraphics[height=1.8in]{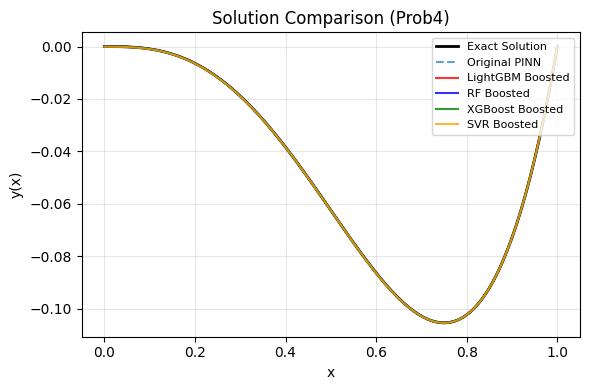}
	\caption{Solution for 70 000 iterations}
	\label{n4v11}
\end{subfigure}%
~ 
\begin{subfigure}[t]{0.5\textwidth}
	\centering
	\includegraphics[height=1.8in]{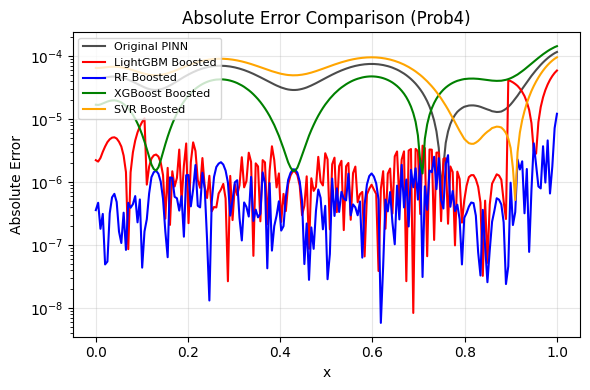}
	\caption{The Absolute Error}
	\label{n4v12}
\end{subfigure}
\caption{The Solutions and Absolute Error for Problem 4}
\label{fig4a}
\end{figure}

\begin{figure}[t!]
	\centering
	\begin{subfigure}[t]{0.45\textwidth}
		\centering
		\includegraphics[height=1.8in]{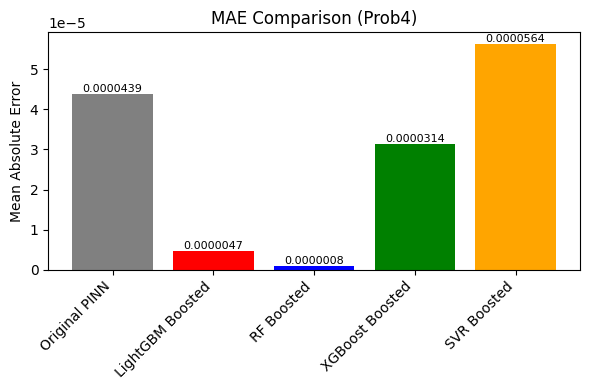}
		\caption{The Mean Absolute Error}
		\label{n4v21}
	\end{subfigure}%
	~ 
	\begin{subfigure}[t]{0.5\textwidth}
		\centering
		\includegraphics[height=1.8in]{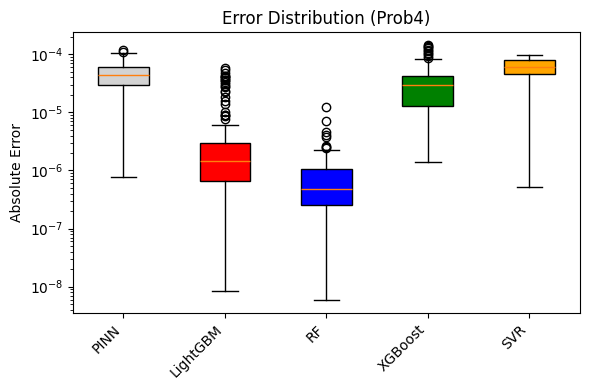}
		\caption{The Absolute Error Distribution}
		\label{n4v22}
	\end{subfigure}
	\caption{The Mean Absolute Error and Absolute Error Distribution for Problem 4}
	\label{fig4b}
\end{figure}

Figures \ref{n4v11} and \ref{n4v12} show the solution and absolute error of problem 4. And Figures \ref{n4v21} and \ref{n4v22} show the MAE and error distribution for the PINN and the boosted PINNs respectively. In Table \ref{comp4}, the absolute errors of PINN is compared with four other methods, in the literature, for solving Problem 4. The PINN demonstrates consistently competitive performance, maintaining stable errors on the order of \(10^{-5}\) across the entire domain, except at $x=1$. While the MLP ANN and the Monte Carlo method \(MC_2\) occasionally achieve lower errors at specific points, the PINN generally outperforms the Chebyshev Neural Network (ChNN) and the other Monte Carlo variant (\(MC_1\)). 

The results in Table \ref{comp4b} compare the absolute errors of a standard PINN with four boosted ML algorithms for Problem 4. In this comparison, the RF-Boosted and LightGBM-Boosted methods consistently achieve the lowest errors, often reducing the error by one to three orders of magnitude compared to the standard PINN - with RF-Boosted reaching errors as low as \(10^{-8}\). The XGBoost-Boosted model also generally outperforms the PINN, though its improvement is less than that of the other tree-based methods. The SVR-Boosted model shows mixed results, sometimes yielding lower errors than the PINN and sometimes having higher errors than the PINN, but frequently performing worse than the boosted tree algorithms. 

\begin{table}[H]
	\centering
\caption{Absolute Error Comparison for Problem 4}
\begin{tabular}{|c|c|c|c|c|c|}
	\hline
	x & PINNs  & MLP ANN \cite{verma2019} & ChNN \cite{mall2015b} &  $MC_1$ \cite{el2023novel}  &  $MC_2$ \cite{el2023novel}  \\
	\hline
	0.1 &  3.32e-05    &   1.1e-05  & 3.1e-04  &  2.49e-05 & 9.86e-06 \\
	\hline
	0.2 &  5.12e-05    &  1.0e-06 &  5.9e-04 & 8.47e-05 &  2.22e-05  \\
	\hline
	0.3 &  6.63e-05    & 3.3e-05  &   3.3e-04 &  1.58e-04  &  1.09e-05 \\
	\hline
	0.4 &  3.18e-05    &  4.8e-05 & 1.2e-05 &  2.22e-04   &  1.50e-05  \\
	\hline	
	0.5 &  4.66e-05    &  2.3e-05 &  6.8e-04 &  2.57e-04  &  3.62e-05  \\
	\hline
	0.6 &  7.51e-05   &  2.7e-05 & 2.8e-05 &  2.41e-04  &  6.62e-06  \\
	\hline
	0.7 &  2.97e-05   &  5.4e-05 &  3.2e-04 &  1.51e-04  &  2.81e-04  \\
	\hline
	0.8 &  1.61e-05    & 1.2e-05 & 2.1e-04 &  3.27e-05  &  3.97e-04  \\
	\hline
	0.9 &  1.89-05   & 8.4e-05  &  1.2e-04 &  3.33e-04  &   6.63e-05 \\
	\hline
	1.0 &  1.16e-04    &  1.0e-05  &  1.1e-05 &  7.70e-04  &  9.47e-04   \\
	\hline
\end{tabular}
\label{comp4}
\end{table}	

\begin{table}[H]
	\centering
	\caption{Absolute Error Comparison of PINNs with the Boosted Algorithms - Problem 4}
	\begin{tabular}{|c|c|c|c|c|c|}
		\hline
		x & PINN  & LightGBM-Boosted & RF-Boosted & XGBoost-Boosted   &  SVR-Boosted \\
		\hline
		0.1 & 3.32e-05 & 9.06e-06 & 4.41e-08 & 5.66e-06	& 5.38e-05 \\ \hline
		0.2 & 5.12e-05 & 2.24e-07 & 1.30e-06 & 2.37e-05	& 7.18e-05 \\ \hline	
		0.3 & 6.63e-05 & 9.21e-07 & 9.96e-07 & 3.87e-05 & 8.69e-05 \\ \hline
		0.4 & 	3.18e-05 & 2.34e-07	& 1.71e-07	& 4.30e-06 & 5.25e-05 \\ \hline
		0.5 &  4.66e-05 & 2.30e-06	& 2.89e-08	& 1.91e-05	& 6.72e-05 \\ \hline	
		0.6 &  7.51e-05 &  7.63e-07	& 1.23e-06	& 4.76e-05 & 9.57e-05 \\ \hline
		0.7 & 2.97e-05 & 3.12e-07	& 2.22e-06	& 2.11e-06 & 5.03e-05 \\ \hline		
		0.8 & 1.61e-05 & 8.76e-07 & 3.25e-07 & 4.37e-05	& 4.49e-06 \\ \hline
		0.9 & 1.89e-05 & 3.84e-05 &	2.09e-07	& 4.65e-05	& 1.67e-06 \\ \hline		    
		1.0 & 1.16e-04 &  5.92e-05 & 1.22e-05 & 1.44e-04	& 9.59e-05 \\ \hline    
	\end{tabular}
	\label{comp4b}
\end{table}

\paragraph{Problem 5:}
The fifth example is
\begin{equation}
	\frac{d^2 y}{d x^2} + \frac{6}{x} \frac{d y}{dx} + 14y = -4y\ln(y)
	\label{prob5}
\end{equation}
with the initial conditions $ y(0) = 1, \frac{d y}{d x}(0) = 0 $. The exact solution is
\begin{equation}
	y(x) = e^{-x^2}.
\end{equation} 

\begin{figure}[t!]
	\centering
	\begin{subfigure}[t]{0.45\textwidth}
		\centering
		\includegraphics[height=1.8in]{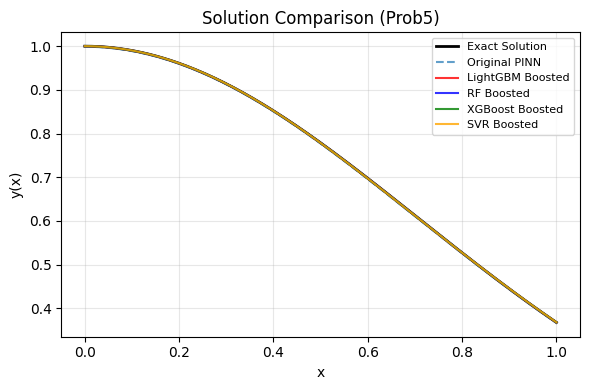}
		\caption{Solution for 50 000 iterations}
		\label{n5v11}
	\end{subfigure}%
	~ 
	\begin{subfigure}[t]{0.5\textwidth}
		\centering
		\includegraphics[height=1.8in]{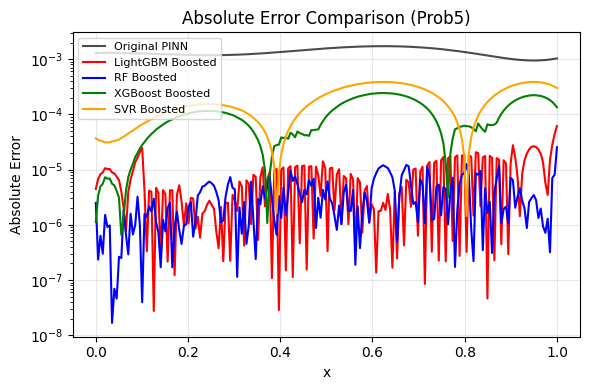}
		\caption{The Absolute Error}
		\label{n5v12}
	\end{subfigure}
	\caption{The Solutions and Absolute Error for Problem 5}
	\label{fig5a}
\end{figure}

\begin{figure}[t!]
	\centering
	\begin{subfigure}[t]{0.45\textwidth}
		\centering
		\includegraphics[height=1.8in]{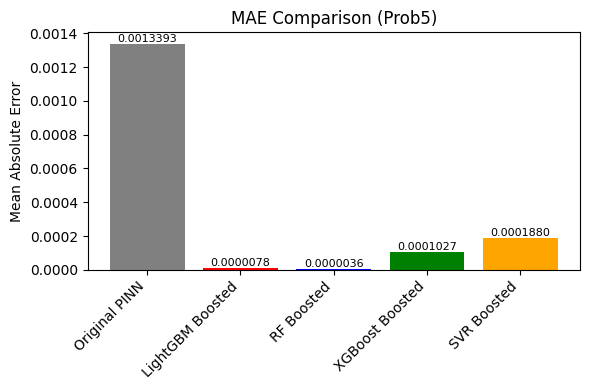}
		\caption{The Mean Absolute Error}
		\label{n5v21}
	\end{subfigure}%
	~ 
	\begin{subfigure}[t]{0.5\textwidth}
		\centering
		\includegraphics[height=1.8in]{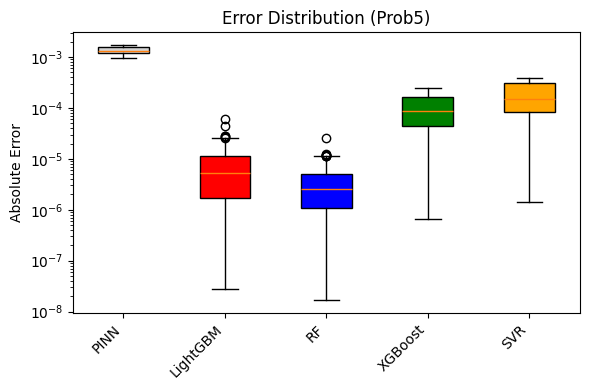}
		\caption{The Absolute Error Distribution}
		\label{n5v22}
	\end{subfigure}
	\caption{The Mean Absolute Error and Absolute Error Distribution for Problem 5}
	\label{fig5}
\end{figure}

Figures \ref{n5v11}, \ref{n5v12}, \ref{n5v21} and \ref{n5v22} show the solution, abute error, MAE and the error distribution for problem 5 respectively. RF still gives the lowest error among the ML models used in boosting the PINN. The results in Table \ref{comp5} compare the absolute errors of PINN with four other methods - ChNN, two Monte Carlo variants (\(MC_1\), \(MC_2\)), and the Legendre Wavelet Method (LWM) - for solving Problem 5. In this comparison, the PINN exhibits the highest error among the methods, consistently around \(10^{-3}\), while the ChNN and LWM achieve significantly lower errors, often as small as \(10^{-6}\) to \(10^{-7}\) at various points. The Monte Carlo methods also perform better than the PINN with errors generally on the order of \(10^{-4}\). This indicates that for Problem 5, the standard PINN is less accurate than the other benchmarked numerical techniques, suggesting that either the problem's characteristics or the specific PINN implementation used here may not be optimally suited for this particular differential equation.

The results in Table \ref{comp5b} compare the absolute errors of the standard PINN with four boosted ML algorithms for Problem 5. In a stark contrast to the PINN, which exhibits relatively high errors on the order of \(10^{-3}\), all boosted methods - especially RF-Boosted and LightGBM-Boosted  - achieve lower errors, often in the range of \(10^{-6}\) to \(10^{-8}\). This represents an improvement of several orders of magnitude over the base PINN. The XGBoost-Boosted and SVR-Boosted models also perform significantly better than the PINN, though they are generally less accurate than the tree-based boosting approaches. This result underscores the benefit of applying ensemble boosting techniques to PINN, as they effectively overcome the limitations of the standard PINN and deliver superior accuracy.

\begin{table}[H]
	\centering
	\caption{Absolute Error Comparison for Problem 5}
	\begin{tabular}{|c|c|c|c|c|c|}
		\hline
		Input data & PINNs  &  ChNN \cite{mall2015b} &  $MC_1$ \cite{el2023novel} & $MC_2$ \cite{el2023novel}  &  LWM \cite{aminikhah2013numerical} \\
		\hline
		0.1 &  1.26e-03  & 1.00e-06  &  1.95e-04  & 8.20e-05  & 4.89e-06  \\
		\hline
		0.2 &  1.18e-03   &  1.00e-05  &  3.78e-04 &  9.49e-05  &  6.84e-06\\
		\hline
		0.3 &  1.19e-03   & 1.00e-06  &  5.33e-04  &  1.35e-04  &  8.03e-07 \\
		\hline
		0.4 &  1.34e-03  &  9.90e-05 &  6.51e-04   &  1.82e-04 &  8.38e-06 \\
		\hline	
		0.5 &   1.57e-03  &  1.00e-04 &  7.25e-04  &  2.12e-04 &  1.28e-05  \\
		\hline
		0.6 &   1.71e-03   &  8.70e-07 &  7.55e-04  &   2.47e-04 &  5.32e-05  \\
		\hline
		0.7 &  1.63e-03    & 1.01e-04 &  7.41e-04 &  2.15e-04  &  2.06e-04 \\
		\hline
		0.8 &  1.32e-03  &  9.80e-07  &  6.88e-04  &  2.24e-04 &  5.93e-04 \\
		\hline
		0.9 &  9.97e-04   &  5.00e-05  &   6.12e-04  &  1.86e-04  &  1.41e-03  \\
		\hline
		1.0 &   1.03e-03  &  5.21e-05  &   5.14e-04  &  1.19e-04  &  3.07e-03 \\
		\hline
	\end{tabular}
	\label{comp5}
\end{table}	

\begin{table}[H]
	\centering
	\caption{Absolute Error Comparison of PINNs with the Boosted Algorithms - Problem 5}
	\begin{tabular}{|c|c|c|c|c|c|}
		\hline
		x & PINN  & LightGBM-Boosted & RF-Boosted & XGBoost-Boosted   &  SVR-Boosted \\
		\hline
		0.1 &  1.26e-03 & 2.47e-05 & 3.96e-08	& 2.81e-05	&	6.61e-05 \\ \hline
		0.2 & 1.18e-03 & 1.06e-06	&	2.14e-06	& 1.04e-04	& 1.42e-04 \\ \hline	
		0.3 & 1.19e-03 &  2.80e-06	&	4.41e-06 &	9.29e-05 &	1.30e-04	\\ \hline
		0.4 & 1.34e-03 & 9.98e-06 & 1.33e-06 & 3.83e-05 & 1.86e-05  \\ \hline
		0.5 & 1.57e-03  & 3.34e-07 & 	6.09e-06 & 9.73e-05	& 2.40e-04 \\ \hline	
		0.6 & 1.71e-03 &  1.92e-06	& 	6.81e-06 & 2.38e-04 & 3.81e-04 \\ \hline
		0.7 & 1.63e-03 & 9.20e-06 & 6.68e-06	& 1.60e-04 & 3.04e-04	 \\ \hline		
		0.8 & 1.32e-03  &  3.24e-07 & 1.26e-05 & 6.10e-05 & 1.45e-06	\\ \hline
		0.9 & 9.97e-04 & 2.76e-05 & 6.45e-06	& 1.68e-04 &	3.32e-04	\\ \hline		    
		1.0 & 1.03e-03 &  6.14e-05 &	2.53e-05 &	1.34e-04 & 2.99e-04 \\ \hline    
	\end{tabular}
	\label{comp5b}
\end{table}

\paragraph{Problem 6:}
The sixth example is given as 
\begin{equation}
	\frac{d^2 y}{d x^2} + \frac{2}{x} \frac{d y}{dx} = - 4 (2 e^y + e^{\frac{y}{2}})
	\label{prob6}
\end{equation}
with the initial conditions $ y(0) = 0,  \frac{d y}{d x}(0) = 0 $. The exact solution is 
\begin{equation}
	y(x) = -2 \ln(1 + x^2)
\end{equation}

\begin{figure}[t!]
	\centering
	\begin{subfigure}[t]{0.45\textwidth}
		\centering
		\includegraphics[height=1.8in]{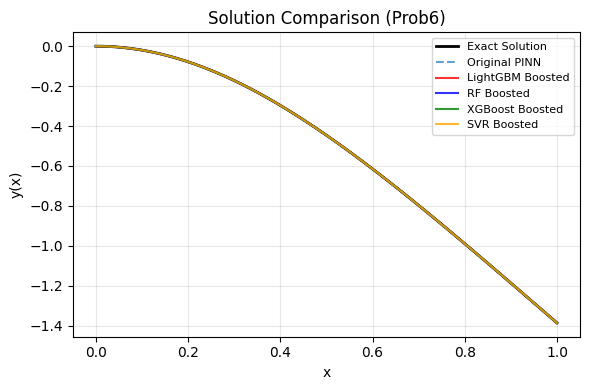}
		\caption{Solution for 30 000 iterations}
		\label{n6v11}
	\end{subfigure}%
	~ 
	\begin{subfigure}[t]{0.5\textwidth}
		\centering
		\includegraphics[height=1.8in]{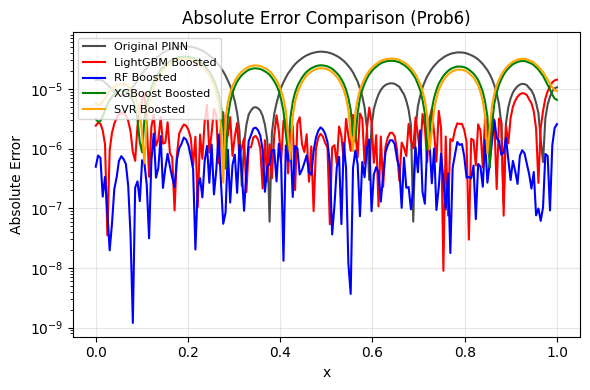}
		\caption{The Absolute Error}
		\label{n6v12}
	\end{subfigure}
	\caption{The Solutions and Absolute Error for Problem 6}
	\label{fig6a}
\end{figure}

\begin{figure}[t!]
	\centering
	\begin{subfigure}[t]{0.45\textwidth}
		\centering
		\includegraphics[height=1.8in]{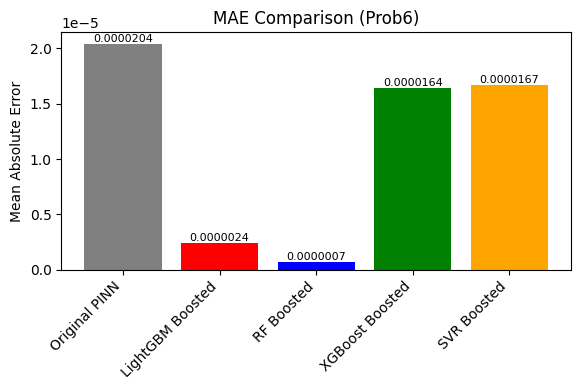}
		\caption{The Mean Absolute Error}
		\label{n6v21}
	\end{subfigure}%
	~ 
	\begin{subfigure}[t]{0.5\textwidth}
		\centering
		\includegraphics[height=1.8in]{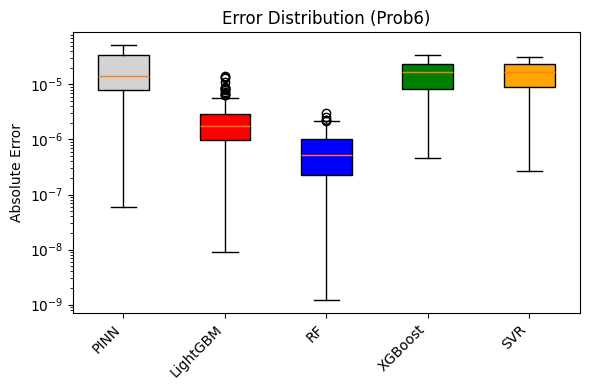}
		\caption{The Absolute Error Distribution}
		\label{n6v22}
	\end{subfigure}
	\caption{The Mean Absolute Error and Absolute Error Distribution for Problem 1}
	\label{fig6b}
\end{figure}

The solution, absolute errors, MAE and error distribution comparison for Problem 6 are given in Figures \ref{n6v11}, \ref{n6v12}, \ref{n6v21} and \ref{n6v22} respectively. Table \ref{comp6} shows the comparison of the absolute errors of PINN with three other numerical methods - two Monte Carlo-based approaches ($MC_1$ and $MC_2$), and Legendre wavelet method (LWM) - across the input data points. Unlike the results in Problem 5, the PINN demonstrates a substantial and consistent advantage, with errors remaining very low (on the order of \(10^{-5}\) to \(10^{-6}\)) across the entire domain. In contrast, both Monte Carlo methods exhibit errors that are one to three orders of magnitude higher (around \(10^{-3}\) to \(10^{-4}\)), and the LWM shows particularly large error growth, especially as \(x\) approaches 1, where its error escalates to \(10^{-2}\).

Additionally, the results in Table \ref{comp6b} compare the absolute errors of the PINN with four boosted ML algorithms for Problem 6. While the PINN already performs well for this problem, the boosted methods - particularly RF-Boosted and LightGBM-Boosted - manage to achieve even lower errors, often in the range of \(10^{-6}\) to \(10^{-7}\). The XGBoost-Boosted and SVR-Boosted models show more variable performance, sometimes reducing errors compared to the PINN but occasionally yielding slightly higher errors at certain points. Overall, this comparison demonstrates that boosting techniques can further refine the accuracy of an already high-performing PINN. 

\begin{table}[H]
	\centering
	\caption{Absolute Error Comparison for Problem 6}
	\begin{tabular}{|c|c|c|c|c|}
		\hline
		Input data & PINNs    &  $MC_1$ \cite{el2023novel} & $MC_2$ \cite{el2023novel} & LWM \cite{aminikhah2013numerical}  \\
		\hline
		0.1 &  1.80e-05   &  3.90e-04 &  2.04e-04  & 1.57e-04 \\
		\hline
		0.2 &   5.15e-05   &  7.58e-04 &  4.59e-04 & 7.73e-05\\
		\hline
		0.3 &   6.23e-06  &  1.07e-03 &  7.26e-04 & 1.26e-04 \\
		\hline
		0.4 &   1.06e-05   &   1.33e-03  &  9.84e-04  & 4.61e-05 \\
		\hline	
		0.5 &   4.10e-05    &  1.53e-03  &  1.21e-03 & 2.03e-04  \\
		\hline
		0.6 &   4.59e-06    &   1.66e-03  &  1.31e-03 & 3.85e-04 \\
		\hline
		0.7 &  7.63e-06   &  1.73e-03 &  1.31e-03  & 1.27e-03 \\
		\hline
		0.8 &   3.95e-05   &  1.75e-03  &  1.24e-03 & 4.38e-03 \\
		\hline
		0.9 &   8.70e-06   &  1.74e-03  &   1.16e-03  & 1.19e-02 \\
		\hline
		1.0 &  1.06e-05  &  1.70e-03 &  1.07e-03 & 2.69e-02  \\
		\hline
	\end{tabular}
	\label{comp6}
\end{table}	

\begin{table}[H]
	\centering
	\caption{Absolute Error Comparison of PINNs with the Boosted Algorithms - Problem 6}
	\begin{tabular}{|c|c|c|c|c|c|}
		\hline
		x & PINN  & LightGBM-Boosted & RF-Boosted & XGBoost-Boosted   &  SVR-Boosted \\
		\hline
		0.1 & 1.80e-05 & 6.39e-06 & 6.39e-07 &  8.64e-07 & 	1.76e-06 \\ \hline
		0.2 & 5.15e-05 &  2.05e-06	& 1.12e-06	&	3.43e-05 & 3.17e-05 \\ \hline	
		0.3 & 6.23e-06 & 1.60e-06 &	7.89e-07  &	1.10e-05  & 1.36e-05 \\ \hline
		0.4 & 1.06e-05 & 6.28e-07 & 5.67e-07 & 	6.56e-06 & 9.16e-06 \\ \hline
		0.5 & 4.10e-05 & 1.04e-06 &	1.28e-06 & 	2.38e-05 & 	2.12e-05 \\ \hline	
		0.6 & 4.59e-06 &  5.78e-07	& 3.68e-07	& 2.18e-05	& 2.44e-05 \\ \hline
		0.7 & 7.63e-06 &  8.16e-07	& 2.01e-06 & 9.54e-06 & 1.22e-05 \\ \hline		
		0.8 & 3.95e-05  &  1.28e-06	&	3.26e-07 &	2.23e-05 &	1.97e-05 \\ \hline
		0.9 & 8.70e-06 & 5.03e-06 &	6.24e-07 &	2.59e-05 &	2.85e-05 \\ \hline		    
		1.0 & 1.06e-05 & 1.43e-05 &	2.57e-06 &	6.56e-06 &	9.19e-06 \\ \hline    
	\end{tabular}
	\label{comp6b}
\end{table}

\section{Conclusion}
\label{secconclusion}
This study conducted a comprehensive performance evaluation of PINNs and their boosted variants against a range of established numerical for solving six distinct differential equation problems - both homogeneous and non-homogeneous Lane-Emden type equations. First, the standard PINN demonstrated robust and often superior performance when compared to traditional methods like Chebyshev Neural Networks (ChNN), Monte Carlo variants ($MC_1$, $MC_2$), and hybrid optimizers (PS-SQP, GA-SQP).  For most problems, the PINN maintained low error magnitudes consistently across the domain and frequently outperforming other methods. However, for other problems, the standard PINN was less effective, highlighting that its performance can be problem-dependent.

Second, and most significantly, the integration of ensemble boosting techniques - specifically LightGBM, Random Forest (RF), and XGBoost - with the PINN framework led to a consistent improvement in accuracy across all six problems. The boosted models, particularly RF-Boosted and LightGBM-Boosted, reduced absolute errors by one to three orders of magnitude compared to the base PINN, achieving error levels as low as $10^{-7}$ to $10^{-8}$. This enhancement was most pronounced in problems where the PINN already performed well, pushing accuracy to new heights, and was crucial also in problems where PINN performed less, effectively compensating for the limitations of the standard PINN.

A primary limitation of this study is its restriction to Lane-Emden type equations - which are ordinary differential equations. The strong performance demonstrated by the boosted PINN framework in this context does not guarantee similar efficacy for partial differential equations (PDEs), which are far more prevalent in scientific modeling and introduce more other challenges.  To address this limitation and extend the impact of the research, future work should prioritize the generalization of the boosted PINN framework to solve PDEs. 

In summary, while PINN alone present a powerful and versatile tool for solving differential equations, their fusion with advanced ensemble machine learning techniques unlocks substantially greater potential. The hybrid boosted PINN approach emerges as a highly promising methodology, combining the physics-driven generalization of PINNs with the superior predictive precision of boosted ensembles to deliver state-of-the-art accuracy and robustness for computational modeling across a diverse set of challenges. This work strongly advocates for the continued development of such hybrid frameworks to tackle increasingly complex scientific and engineering problems.

\bibliographystyle{unsrt}
\bibliography{reference2.bib}

\end{document}